# Rejuvenating the structure and rheological properties of silica nanocomposites based on natural rubber

Kanyarat Boonsomwong,[1,2] Anne-Caroline Genix,[2] Edouard Chauveau,[2] Jean-Marc Fromental,[2] Philippe Dieudonné-George,[2] Chakrit Sirisinha,[3] and Julian Oberdisse [2]

[1] *Department of Chemistry, Faculty of Science, Mahidol University, Rama VI Road, Phayathai, 10400 Bangkok, Thailand*

[2] *Laboratoire Charles Coulomb (L2C), Université de Montpellier, CNRS, F-34095 Montpellier, France*

[3] *Rubber Technology Research Centre, Faculty of Science, Mahidol University, Salaya Campus, Phutthamonthon IV Road, Salaya, 73170 Nakhonpathom, Thailand*

**ABSTRACT.**

The antagonistic effect of processing and thermal annealing on both the filler structure and the polymer matrix is explored in polymer nanocomposites based on natural rubber with precipitated silica incorporated by coagulation from aqueous suspension followed by roll-milling. Their structure and linear and non-linear rheology have been studied, with a particular emphasis on the effect of high temperature thermal treatment and the number of milling passes. Small-angle X-ray scattering intensities show that the silica is organized in small, unbreakable aggregates containing ca. 50 primary nanoparticles, which are reorganized on a larger scale in filler networks percolating at the highest silica contents. As expected, the filler network structure is found to be sensitive to milling, more milling inducing rupture, as evidenced by the decreasing Payne effect. After thermal treatment, the nanocomposite structure is found to be rejuvenated, erasing the effect of the previous milling on the low-strain modulus. In parallel, the dynamics of the samples described by the rheology or the calorimetric glass-transition temperature remain unchanged, whereas the natural latex polymer network structure is modified by milling towards a more fluid-like rheology, and cannot be recovered.

## 1. INTRODUCTION

Polymer nanocomposites (PNCs) are made of hard nanoparticles (NPs) embedded in a polymer matrix, with the general aim to improve the rheological properties of the crosslinked or uncrosslinked polymer for applications. [1-3] This so-called reinforcement effect has many different facets according to the property to be optimized: the addition of filler NPs increases the low-strain modulus, [4, 5] usually accompanied by strain-softening known as the Payne effect. [6] Moreover, resistance to rupture, i.e. toughness and sometimes recovery, is generally improved. Most of these effects are related to two phenomena: (i) the modification of the dynamics of the polymer molecules close to a hard interface, possibly aided by chemical coupling and resulting in molecular mobility or glass-transition temperature ($T_g$) changes, and (ii) the dispersion of the NPs in the matrix, which can evolve from percolated filler networks, [7, 8] to aggregates, [9, 10] or individually dispersed NPs . [11]

Control of the type of dispersion can be obtained by tuning interactions between filler NPs, [12, 13] between the filler and the matrix via a coupling agent, [14] or via processing [15, 16] and in particular melt blending. Solid-state mixing and milling are intended to avoid the use of solvent during nanocomposite processing. They allow breaking precipitated silica agglomerates into smaller aggregates and distribute them in the hydrophobic polymer matrix. The impact of the milling conditions (time, temperature, orientation) has been intensely studied – mostly in the fifties – and found to affect the filler dispersion state as seen by electron microscopy, [17] the bound rubber content, [18] and various mechanical properties [19, 20] of vulcanized rubbers. Nanocomposite formation from mixing polymer latex and silica suspensions is the approach chosen here to incorporate the silica filler, followed by coagulation and, once dried, milling. The process of latex film formation has been deeply investigated in the literature for synthetic [21, 22] and natural rubber (NR) [23] systems. In all cases, the process starts with the latex beads coming in close contact upon drying. Cellular latex films have been described in the literature for synthetic nanolattices, [21, 24] and due to monodispersity and order they possess rheological properties reminiscent of foams. Such synthetic latex structures may thus generate an initial scaffold for the silica, which cannot penetrate inside the latex beads and thus concentrates in the interstitial sites, possibly in the form of agglomerates. A similar approach has been investigated by Peng et al where NR latex beads were vulcanized beforehand, demonstrating a higher thermal resistance. [25] Also for NR, it was shown by AFM studies monitoring the surface morphology that there is no well-defined ordering of the beads (due to strong polydispersity in size) but a randomly packed structure with ill-defined contours. [26] Due to the low glass-transition temperature, $T_g$ (about -65°C, i.e. NR latex beads are liquid droplets as shown in the SI), chains may then interdiffuse until the initial organization in the form of beads has completely disappeared, and a molecularly dispersed melt of polymer chains is obtained.

After drying the coagulated mixture of silica and latex, milling is known to contribute also to homogenization of the matrix with possible synthetic additives like processing oils, or naturally present non-isoprene compounds [27, 28] like phospholipids or proteins (about 5%wt), which will be relevant for the present work on natural rubber. These non-isoprene molecules have been recognized to form initially an adsorbed shell surrounding the latex particles in aqueous suspension. In the final uncrosslinked NR matrix, such molecules may self-assemble and induce additional (physical and reversible) crosslinks between the polymer molecules, [29] thereby solidifying the entangled matrix [30] and increasing slightly the rheological response. [31] Pure NR matrices have only little interaction with bare silica particles, [32] favoring a dominant response of the filler network. This is visible in particular around the filler percolation threshold giving rise to a strong increase of the storage modulus.

The dynamical properties of PNCs have been probed by different techniques (like broadband dielectric spectroscopy, [33, 34] NMR, [7, 35] QENS [36, 37] …) depending on the time scale of interest. Rheology is particularly well-suited as it focuses on viscoelastic properties useful for applications. Here both linear (frequency-dependent moduli) and non-linear (Payne effect) rheology will be used to study a system close to industrial applications.

Mechanical rejuvenation of glassy polymers, i.e., submitting the material to large deformation to erase or reverse physical aging has been introduced by Struik. [38, 39] During aging, the free volume decreases, and the dynamics slows down accordingly. The concept of mechanical rejuvenation is still a matter of debate. [40] It has been suggested by McKenna [41] that the application of large stress moves the system back to a different thermodynamic state than the one of the unaged glass. Similarly to aging of glassy polymers, low intensity mechanical excitations have been reported to allow reconfiguration of glassy states in colloids counter-balancing aging phenomena. [42-44] In both cases, energy input thus moves the system back to younger states.

Another way to input energy into the system is a high-temperature treatment. In polymer nanocomposites, the general impact of thermal annealing is to move the system closer to thermodynamic equilibrium, by allowing the exploration of a wider range of particle configurations, and polymer conformations. Due to the lower viscosity, above-mentioned additives and polymer may be homogenized, and internal stresses may be relaxed. Simultaneous, the NPs acquire higher mobility, and because of the usually present intrinsic incompatibility between polymer and NPs, the latter tend to aggregate during annealing. [45, 46] Annealing thus cancels processing history and this effect may also be termed rejuvenation, in the sense that it moves the system back to its initial, aggregated state.

It is the objective of the present paper to study the influence of annealing and processing on both the filler structure and the polymer. We report on the progressive destruction of both the natural crosslinking of the polymer matrix and the silica network by milling as a function of the filler content, and demonstrate the possibility of re-assembly or rejuvenation of the silica structure using thermal treatment. In order to highlight the effect of reorganization of the silica network (via silica-silica interactions), no polymer-silica coupling agents have been introduced, and volume fractions up to percolation have been chosen. In absence of chemical coupling of the filler to the network, attractive interparticle interactions induce the formation of aggregates containing some embedded polymer (usually called occluded rubber [47]), which leads to an increase of the effective filler volume fraction and may generate hard percolating paths at rather low nominal filler volume fractions, thereby reinforcing the modulus. Upon milling, both silica aggregates or networks and the matrix structure containing physical crosslinks due to the non-isoprene compounds may be destroyed, with a strong impact on the rheological properties of the PNCs. Thermal treatment will be shown to promote the regeneration of the silica network, generating stiffening, whereas the non-isoprene matrix compounds are better dispersed, the corresponding crosslinks broken up at high temperature, inducing a more fluid-like rheological behavior.

## 2. MATERIALS AND METHODS

**Sample Formulation.** The natural rubber (NR, from Hevea Brasiliensis, Thai Rubber Latex Co. Ltd., Chonburi, Thailand) has a typical molecular mass of ca. 1 000 kg mol$^{-1}$, and polydispersity above 2. It was used as received, in the form of a 60%wt aqueous suspension with high ammonia grade (0.7%wt). The latex bead size is approximately one micron as deduced from optical microscopy (see SI). The precipitated silica powder (MBJ Enterprise Co. Ltd.) has a specific surface of 185 m$^2$/g, which is compatible with primary beads described in the scattering analysis of radius 9 nm and polydispersity ca. 22% (see SI for details). The silica was processed in the form of an aqueous suspension (20%wt).

The silica filled-rubber PNCs were prepared by using a latex compounding method without any curing or coupling agents. The filler content was varied from 10 to 50 phr by adjusting the quantity of silica suspension as indicated in Table 1. In each case, a fixed amount of compounding agents was used: for 100 g of latex (corresponding to 166.7 g of suspension at 60%wt), 13 g of naphthenic oil (rubber processing oil, PSP Specialties Co. Ltd.), 3 g of stearic acid (Kij Paiboon Chemical Ltd.), and 1 g of antioxidant (butylated hydroxytoluene, BHT) were added. Moreover, the unfilled compound was prepared for comparison.

| Samples | Unfilled | 10 phr | 30 phr | 40 phr | 50 phr |
|---|---|---|---|---|---|
| 20%wt silica suspension (g) | 0 | 50 | 150 | 200 | 250 |
| $\Phi_{NP}$ (%vol) | - | 3.4 | 8.6 | 11.2 | 12.7 |

**Table 1.** Formulation of unfilled and silica filled-rubber PNCs loaded with 10, 30, 40 and 50 phr of silica and silica volume fraction $\Phi_{NP}$ deduced from TGA.

The compounding was done with the following protocol: stearic acid and BHT were molten in napthenic oil at 80°C. Then, the oil was emulsified with 0.15%wt of surfactant (potassium oleate, Lucky Four Co. Ltd.) for 200 ml at the same temperature for 10 minutes. The emulsion was cooled down for 5 minutes and the rubber latex was added under stirring for 10 minutes. Subsequently, the 20%wt silica suspension was added to the preceding mixture, and the formed blend was coagulated by incorporation of a 5%wt $CaCl_2$ solution. The coagulated compounds were extracted as sheets with a rubber crepe machine and dried at room temperature for 48 h. To gain uniformity, the dried compounds were milled using a two-roll mill (about 30 passes, gap 0.65 mm) at 50°C and kept to dry at room temperature for another 24 h. Then, those compounds were re-milled at 50°C with various numbers of milling passes at 0.1 mm. Four different numbers of milling passes, 10, 20, 30 and 40, were carried out in order to investigate the effect of milling pass on the silica dispersion. Note that the same treatment was applied to the unfilled rubber matrix for reference.

The appearance of unfilled and silica-filled samples is shown in Figure 1. Their color slightly evolves becoming more brownish upon addition of silica. In parallel, the samples become more translucent with higher number of milling passes.

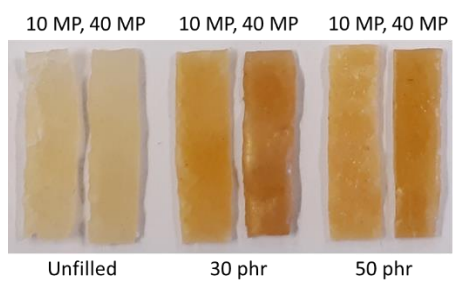

**Figure 1.** Pictures of unfilled compounds and PNCs (30 and 50 phr of silica) with 10 or 40 milling passes. Approximate sample size is 5 mm x 25 mm.

**Latex film formation.** The glass-transition temperature of the NR being well below ambient temperature, the NR latex beads are expected to be emulsified liquid droplets, allowing an easy film formation by coalescence. Optical microscopy pictures of drops of latex suspension are shown in the SI. From the corresponding size distribution, we found that the latex beads are typically 500 nm in diameter. We also report in SI the pictures obtained after drying these drops at room temperature during a couple of hours. The drops coalesce and the latex morphology changes with a significant shift (x10) of the effective size distribution. Due to the attractive Van der Waals interactions between silica NPs, and the absence of favorable polymer-NP interactions (no coupling agents), aggregation of the silica is to be expected, with aggregates embedded in the NR matrix generated by latex coalescence.

**Thermogravimetric analysis (TGA).** The silica fractions in PNCs were measured by TGA (TGA2, Mettler Toledo) using a first ramp at 10 K/min from 25 to 500°C under nitrogen, followed by a second ramp at 20 K/min under air up to 900°C. The weight fractions of silica were associated with the weight residue at 900°C. The volume fractions were determined by mass conservation using the densities of the natural rubber and silica, 0.93 and 2.2 g.cm$^{-3}$, respectively. They are given in Table 1. The standard deviation as determined by measuring 4 pieces of the same sample (30 phr PNC) was found to be 0.1%.

**Differential Scanning Calorimetry (DSC).** The glass-transition temperature was measured by DSC (Q2000, TA instrument, USA) in the modulated mode. Samples of about 10 mg were sealed in aluminum pans and heated at an average rate of 3 K/min (temperature modulation amplitude 0.5 K, period 60 s). $T_g$ was defined as the inflexion point temperature in the reversible heat flow of the second run upon heating. For the PNC samples, $T_g$ shows no variation as compared to the unfilled compounds: $T_g$ = –64.3 ± 0.3 °C for all the silica contents investigated here. The $T_g$ of the raw material prepared without rubber processing oil is found to be higher (-62.5 °C) evidencing the plasticization effect of the oil in the rubber compounds.

**Rheology.** The rheological response in the linear regime of the matrices and nanocomposites (Table 1) was measured using a stress-controlled rheometer AR 2000 (TA Instruments, USA) used in the strain-controlled mode with a plate-plate geometry (diameter 20 mm). To ensure stability of the measurements, an equilibration time of 10 h at 100°C was applied to the PNC samples, as determined independently beforehand. Isothermal frequency sweeps were performed from 100 to 0.02 rad/s at fixed low-deformation level, γ = 0.1%, in the linear regime. The normal force was kept constant to (2 ± 0.1) N. The temperature range from 100°C to 20°C with 10°C step (waiting time 10 min) was investigated. Using the principle of time-temperature superposition, master curves of the storage modulus, G'(ω), and the loss modulus, G"(ω), corresponding to measurements at a reference temperature of 70°C were established. Master curves were built by applying the same shift factors – $a_T$ (horizontal) and $b_T$ (vertical) – to both G' and G''. Note that it was necessary to introduce vertical shifting to achieve a good overlap of the frequency sweeps for all samples, including the ones without silica. Such a temperature correction to the viscoelastic modulus is due to the effect of density that is weakly T-dependent. [48] It follows that $b_T$ values vary much less than their horizontal counterpart, and we found consistently that they do not evolve neither with the silica fraction nor with the milling process.

Non-linear rheology was studied using a Rubber Process Analyzer (Alpha Technologies RPA2000, USA). Dynamic strain sweeps were performed at 70°C with a frequency of 1 Hz and under the same conditions after a temperature isotherm of 5 minutes at 150°C. The latter temperature corresponds to the curing temperature for the preparation of the final products (vulcanized PNCs), which are not studied in the framework of this study but may serve later as comparisons. The amplitude of the Payne effect describing the softening of the storage modulus under strain, ΔG, was obtained from the difference of the storage modulus at low and high strain:

$$\Delta G = G'_{low} - G'_{high} \qquad (1)$$

where the index 'low', and 'high' correspond to γ = 0.56% and 300% of strain amplitude, respectively. Each set of measurements was performed twice to check reproducibility, and we present the average result in the following.

**Small-angle X-ray scattering (SAXS).** SAXS experiments were performed with an in-house setup of the Laboratoire Charles Coulomb, "Réseau X et gamma", Université de Montpellier (France). A high brightness low power X-ray tube, coupled with aspheric multilayer optic (GeniX3D from Xenocs) was employed. It delivers an ultralow divergent beam (0.5 mrad) with a wavelength of 1.5418 Å (8 keV). Scatterless slits were used to give a clean 0.6 mm beam diameter with a flux of 35 Mphotons/s at the sample. We worked in a transmission configuration and scattered intensity was measured by a 2D "Pilatus" 300K pixel detector by Dectris, at a distance of 1.843 m from the sample yielding a q range from $5 \times 10^{-3}$ to 0.2 Å$^{-1}$. The scattering cross section per unit sample volume dΣ/dΩ (in cm$^{-1}$) termed scattered intensity I(q), was obtained by using standard procedures including background subtraction and calibration. [49] The scattering of the unfilled compounds was measured independently to subtract the matrix contribution for each PNC sample.

## 3. RESULTS AND DISCUSSIONS

### 3.1 Structure and rheology of native compounds

**Payne effect.** In pure polymer networks, strain-dependent non-linear rheological properties are commonly observed at rather high strains. On the contrary, the hysteretic softening of filled polymer networks is found at low oscillatory strains, and it is usually termed the Payne effect. [6] This apparition of a non-linear regime of strain-dependent moduli is thought to be predominantly due to breaking of the filler network, as described by the Kraus model. [50] Non-linear strain sweeps measured at 70°C on the PNCs containing 30 phr of silica after different numbers of milling passes (MP) are presented in Figure 2a. These native compounds have not undergone any high-temperature thermal treatment that will be tested in the next section. The strain-induced softening of the elastic modulus is well visible and it is found to strongly depend on the milling conditions: the magnitude of the Payne effect decreases for higher numbers of milling passes. Moreover, the range of the linear domain extends to higher strain with milling. Both observations are in line with a better dispersion of the silica NPs, which tends to break up the underlying filler network. The variation in modulus from low to high strain, $\Delta G$, as defined in the Method sections is summarized in Figure 2b (full symbols). Besides weakening of the Payne effect upon milling, it is observed that it is more pronounced for increasing silica fraction as commonly observed in many PNC systems, [51] whereas the unfilled compounds exhibit virtually no softening of the storage modulus. One may wonder if some of the physical crosslinks of the latex network are also affected by the strong deformations, but as the softening of the unfilled compound is negligible (and independent of MP) as shown in Figure 2b, one may conclude that the silica network dominates the rheological response.

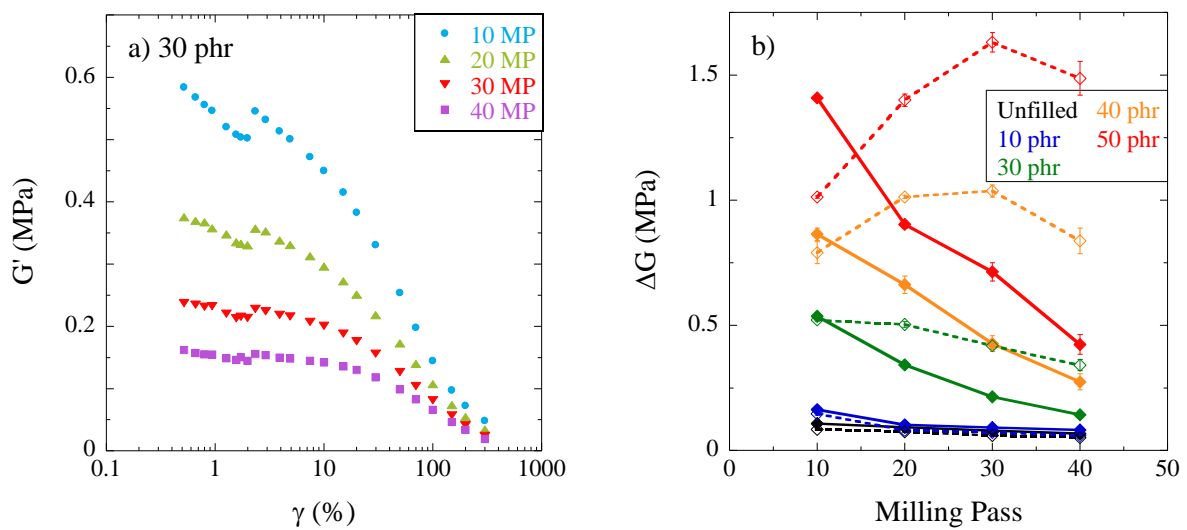

**Figure 2. a)** Strain sweeps of silica PNCs filled with 30 phr after different numbers of milling passes: 10, 20, 30 and 40, T = 70°C and frequency = 1 Hz. **b)** Payne effect of unfilled compounds and PNCs with 10, 30, 40 and 50 phr of silica loading at 70°C versus the number of milling passes (full symbols). The same measurement after a thermal treatment of 5 minutes at 150°C is shown for comparison (empty symbols). Lines are guides to the eye.

**Silica microstructure by SAXS.** The microstructure of the samples in Table 1 in absence of thermal treatment has been studied by small-angle X-ray scattering. First, the SAXS intensities of the unfilled compounds do not show any clear signature of the primary latex bead structure in the investigated q-range, and there is no detectable impact of the number of milling passes (see SI): as discussed in the Methods section, the pure NR latex matrix is thus homogeneous on the 100-nanometer scale probed by SAXS. In Figure 3a, the scattering curves for the series in silica volume fractions in the NR matrix are shown. We used the reduced representation $I(q)/\Phi_{NP}$ with $\Phi_{NP}$ given in Table 1 to compare PNCs of different silica contents and reveal variations in the apparent (due to aggregate polydispersity) inter-aggregate structure factor, $S_{inter}(q)$, which is a characteristic of the spatial arrangements of the silica NPs. The low-q limit of $S_{inter}(q)$ describes the isothermal compressibility and thus the strength of the interactions between aggregates. Its evolution with silica concentration and milling is plotted in the inset of Figure 3a, and will be discussed below.

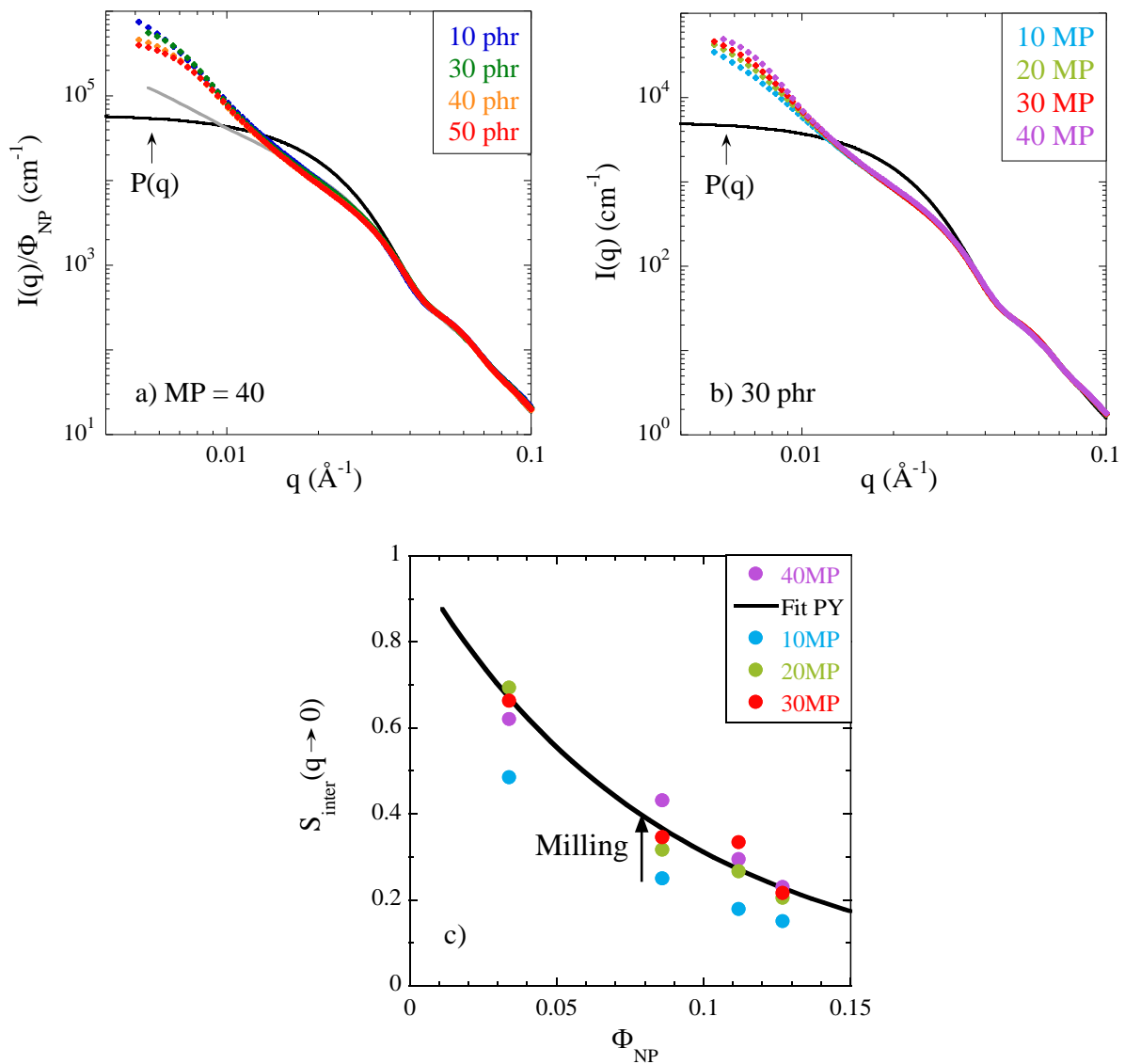

**Figure 3.** SAXS scattered intensity for PNCs with **a)** different silica loadings, all measured after 40 milling passes, and **b)** 30 phr after different numbers of milling passes. The black line is the theoretical form factor of the primary silica nanoparticles. The grey line is the scattered intensity for the silica powder. **c)** Low-q limit of the inter-aggregate structure factor $S_{inter}(q \to 0)$ as a function of $\Phi_{NP}$ for different MPs as indicated in the legend. The black line is the Percus-Yevick prediction for such aggregates.

The absence of evolution of the renormalized intensity $I/\Phi_{NP}$ in Figure 3a at intermediate and high-q contains two different pieces of information. Naturally, the same silica NP is highlighted by the scattering at the highest q where NP surface scattering (Porod law, cf. [49]) is probed. At intermediate q, around q = 0.02 Å$^{-1}$, the scattering of the assembly of NPs is below the theoretical scattering of the form factor of the primary particles indicated by the black curve. This provides information on the repulsive interactions impeding interpenetration between the hard particles, corresponding to a minimum center-to-center distance of (2π/0.03) Å = 20 nm, i.e., ca. two particle radii. Considering that aggregates are made of primary NPs, this corresponds to the close-contact distance. Moreover, the extend of the deviation between the curves (termed the correlation hole due to the dip of I(q) below P(q)) is related to the local volume fraction of particles that we term the internal aggregate density or compacity, κ. [52] One can see a good superposition in this q-range for all PNCs with different silica contents and the original silica powder (rescaled in intensity) in Figure 3a, and for different milling in Figure 3b. In all these PNCs, the local arrangement of NPs is thus unchanged by milling or silica content, and has been transferred without modification from the silica powder. The SAXS data in this range thus describe the unmodified average aggregate form factor. In other words, aggregates are considered as similar from one sample to the other and they constitute the unbreakable building units of a larger length-scale silica arrangement as previously observed in styrene-butadiene PNCs. [8]

At the lowest angles, finally, the intensities are found to depend on both milling and silica volume fraction in Figure 3. In an effect similar to the formation of the correlation hole at intermediate angles, the higher silica content and thus higher concentration of aggregates naturally induce a decrease of the intensity described by $S_{inter}(q)$. On the scale probed here (5x10$^{-3}$ Å$^{-1}$), this means that the silica concentration effect is handed down to ca. 100 nm, presumably by forming networks of repulsively interacting aggregates (hard sphere [53, 54]), whereas the local silica aggregate structure observed at intermediate angles (corresponding to 20 nm) remains unchanged. The effect of the number of milling passes, finally, shows that the silica network structure of repulsively interacting aggregates is broken up, inducing better dispersion, and thus less repulsive low-q interactions.

For the PNC series obtained after maximum milling (40 MP, Figure 3a), the limiting value of the inter-aggregate structure factor has been determined from the low-q intensity in the following way. First of all, the high- and intermediate-q data can be described by an aggregate form factor, $P_{agg}(q)$, which can be built by a product of interparticle structure factor S(q) and NP form factor P(q), [55] followed by averaging over all aggregates. The low-q limit of $P_{agg}$ gives the aggregate mass by means of the ratio of the second and first moments of the distribution in aggregation number, $<N_{agg}^2>/<N_{agg}>$. The distribution in $N_{agg}$ is directly related to the distribution in aggregate size via the compacity, $N_{agg}$ = $4\pi R_{agg}^3 \kappa /(3V_{NP})$ where $R_{agg}$ is the aggregate radius and $V_{NP}$ the average silica bead volume. We have obtained a first estimate of the average aggregate mass and size via a Guinier analysis of the low-q data of the lowest concentration sample (10 phr and 40 MP). We have then iteratively searched and found a self-consistent solution for I(q→0) by calculating the corresponding value of $S_{inter}$(q→0) using interactions of hard polydisperse aggregates, which we know from extensive Monte Carlo simulations. [8, 52] To put it simple, the first estimate of the aggregate mass and size leads a total aggregate volume (including polymer), which determines hard-sphere interactions, and thus generates a structure factor which is taken into account in a second iteration, slightly adjusting the first guess of aggregate properties in order to agree with the experimental I(q→0). The result is thus a self-consistent description of (unchanged) aggregates of the following average aggregation number: $<N_{agg}>$ = 18, compacity: $\kappa$ = 37%, and polydispersity in size: $\sigma$ = 34%. Another measure of aggregation number is given by $<N_{agg}^2>/<N_{agg}>$ = 52. The corresponding data points for interaction $S_{inter}$(q→0) have been plotted in Figure 3c. They are found to decrease as expected with silica concentration, and by construction they are found to be only slightly scattered around the theoretical black curve describing hard aggregate interactions. [52]

For milling with 10, 20 and 30 MP (Figure 3b), the apparent $S_{inter}$(q→0) can be found by simply dividing the experimental low-q intensity by the now known aggregate form factor $P_{agg}$(q→0). Note that we have analyzed the 40 MP series first based on the hard aggregate interactions, because it is expected to correspond to the most random dispersion. With a short milling (10 MP), the resulting data points for $S_{inter}$(q→0) are found to move down, as with increasing volume fraction: while $\Phi_{NP}$ thus leads to stronger interactions between aggregates, milling is found to weaken them. For long milling times, the interactions saturate and tend to the hard-sphere case.

To summarize structure and rheology, NR PNCs directly after milling and without any thermal treatment are strongly affected by the amount of milling passes: the higher the number of MP, the weaker the Payne effect (i.e., the effect of additional strong deformation), and the better the dispersion of the silica aggregates in the matrix, suggesting rupture of the silica network. SAXS also demonstrates that small unbreakable aggregates are the building units of the silica filler network.

### 3.2 Rejuvenation of silica-NR nanocomposites

Due to the thermodynamic driving force of the suspension of silica NPs in a NR matrix in absence of any coupling agents – which would favor attraction between silica and the polymer –, the silica NPs tend to aggregate and form networks. By applying large deformations to these samples, their internal filler networks have been found to break up (Payne effect) in the preceding section, and the large-scale structure on the scale of unbreakable aggregates is reorganized correspondingly (SAXS). By heating the samples to high temperatures, the viscosity of the polymer matrix is lowered, enhancing the mobility [9, 46] of the NPs in the matrix. The effect of such a treatment is now investigated.

**Silica microstructure after annealing.** The effect of a short but intense (150°C, 5 min) thermal treatment on the microstructure is presented in Figure 4a. As opposed to milling, which leads to a low-q increase of the scattered intensity, a decrease is systematically observed, which we identify as a rejuvenation effect. Using the same analysis as in the previous section before annealing, and in particular the same aggregate form factor, the apparent structure factor at low q can be determined. The corresponding values are plotted in Figure 4b, where they are also compared to the Percus-Yevick prediction of hard polydisperse aggregates [52, 53]. Note that the latter are supposed to remain unchanged in all experiments as shown by the superimposing mid- and high-q intensities.

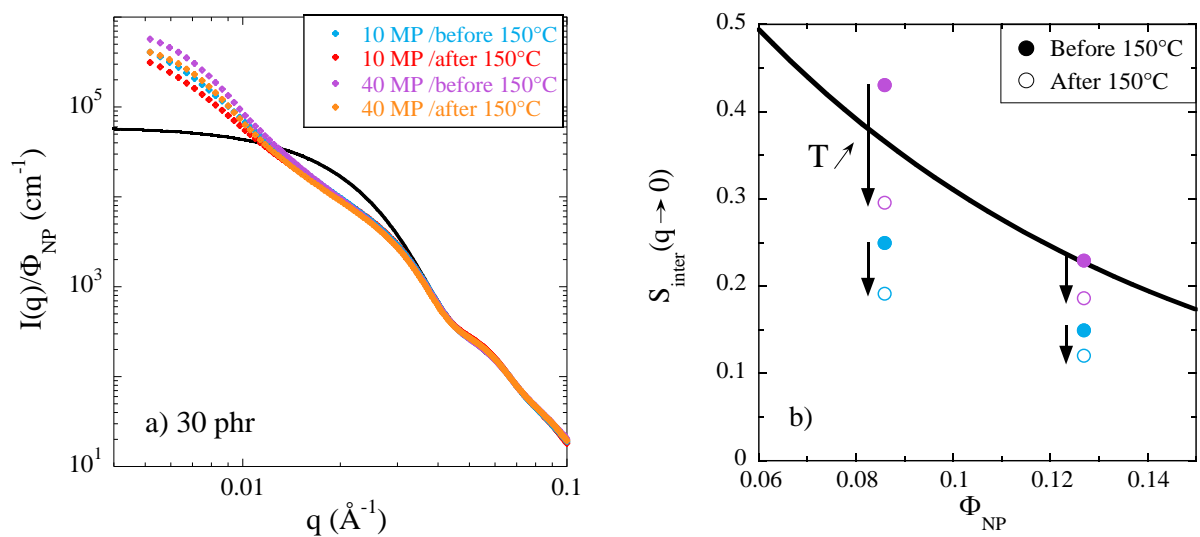

**Figure 4:** Effect of annealing on SAXS scattered intensity for PNCs at 30 phr **a)** for 10 and 40 milling passes before and after annealing as indicated in the legend. The black line is the theoretical form factor of the primary silica nanoparticles. **(b)** Evolution of the low-q limit of the inter-aggregate structure factor $S_{inter}(q \rightarrow 0)$ as a function of $\Phi_{NP}$ with annealing for the two MPs. The black line is the Percus-Yevick prediction for such aggregates.

The important result of Figure 4b is that the apparent structure factor deviates from the ideally random hard aggregate structure illustrated by the polydisperse Percus-Yevick prediction. The low-q structure factor decreases with annealing, returning towards its initial, non-random state, i.e., the sample is indeed rejuvenated. This is also suggested by the arrows in Figure 4b, which point in the opposite direction as in Figure 3c. We will now investigate the impact of this effect on the macroscopic properties of the PNCs.

**Payne effect after rejuvenation.** We study now the impact on mechanical properties of the recovery of the silica network structure under such conditions, and illustrate this network formation by revisiting the Payne effect, which is completed by a study of the linear rheology.

The Payne effect has also been characterized after a thermal treatment of the PNCs. It is well-known that the magnitude of the Payne effect diminishes as the temperature is increased. [56] Here, we looked at the impact of a rest time of 5 minutes at 150°C, prior to measurement at the same temperature of 70°C. The results are shown in Figure 2b where they are compared to the direct measurements at 70°C discussed above. One can see that thermal treatment has a strong impact at the highest filler content, suggesting reorganization of the filler network, whereas at lower concentrations, the ΔG variation approaches the one in absence of heating. In any case, the impact of milling (before thermal treatment) has been cancelled, as one can see from the change in shape of the curves: instead of a continuous decrease with milling as observed before, they are now (mostly) constant or display a lower amplitude of variation, or even increase. In any event, the strength of the Payne effect, ΔG, is comparable after 40 MP and thermal treatment to the one after 10 MP without thermal treatment. It is thus concluded that the short thermal treatment at high temperature (150°C) is sufficient to recreate the filler network which has been broken up by the milling process. This finding agrees nicely with the above SAXS analysis, and will be further backed up by linear rheology in the next section. In particular, the evolution of the modulus with the duration of the thermal treatment allows following such a process.

**Linear rheology after rejuvenation.** PNC samples after milling have undergone a thermal treatment as indicated in the Methods section (100°C, 10h), initially in order to obtain stable rheology curves. A surprising evolution of the linear moduli during this annealing step was found, and finally motivated this article. As indicated in the introduction, two independent network contributions are present in these NR latex-silica systems: the silica, and the polymer network. The impact of annealing on both is opposite (see SI for data monitoring the evolution of the elastic modulus with time, Figure S4): the silica network is reconstituted after having been broken by milling, whereas thermal treatment decreases the contribution of physical crosslinks to the polymer network, presumably by promoting the dispersion of non-polyisoprene compounds as discussed below. After this initial annealing step, the effect of milling on the linear rheology has been studied. An example of master curves (at 70°C) is presented in Figure 5 for a 30 phr nanocomposite having undergone different milling before thermal treatment. All curves have been constructed using vertical and horizontal shift factors as explained in the Methods section. The horizontal shift factors may be used to analyze the evolution of the dynamics with temperature. This result has been reported in the SI, where it is shown that the dynamics can be described by a Vogel-Fulcher-Tammann equation. The dynamics is found to be unaffected by both the silica content – superimposing with the unfilled sample –, and the number of milling passes, in agreement with the absence of evolution of the calorimetric glass-transition temperature for all samples.

The analysis of the shape of the master curves presented in Figure 5 reveals that, at the highest frequency, a common limiting modulus independent of the amount of milling is reached. Our interpretation of this result is that the filler network initially broken up by milling (see Figure 2b) has been recreated by the thermal treatment (rejuvenation) and is probed in this experiment, independently of the initially undergone milling.

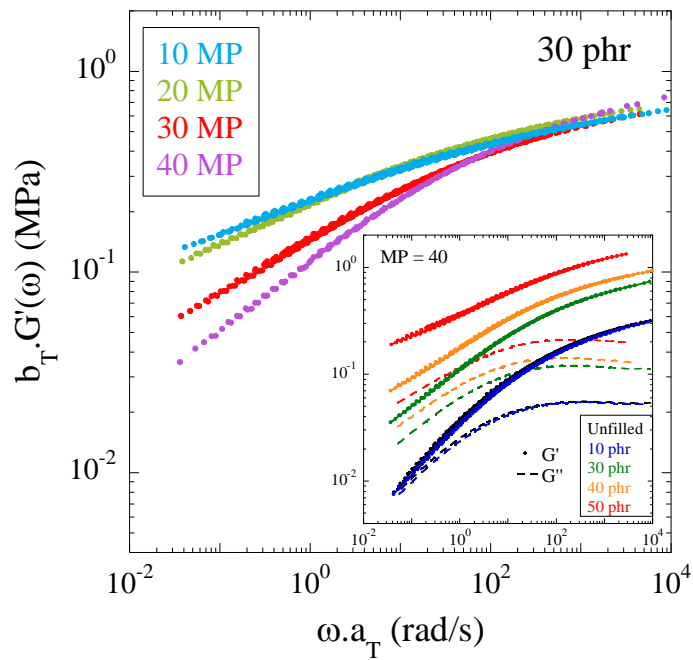

**Figure 5.** Shear moduli G'(ω) using time-temperature superposition as a function of angular frequency at the reference temperature of 70°C of 30 phr PNCs after different number of milling passes. Inset: G'(ω) and G''(ω) for the unfilled compound and PNCs with different silica loadings, all measured after 40 milling passes.

On the other hand, G'(ω) is observed to evolve in the low frequency range, towards a steeper slope, which is indicative of more fluid samples. There is a systematic evolution from solid-like to liquid behavior when increasing the number of milling passes at fixed silica content in Figure 5. A strong deviation from Maxwellian behavior (i.e., scaling laws of $G' \sim \omega^2$ and $G'' \sim \omega^1$ in the terminal regime) is observed. In order to study the reinforcement effect by the filler network, the master curves of the samples containing from 0 up to 50 phr of silica after a milling process of 40 passes are shown in the inset of Figure 5. The presence of nanoparticles leads to a fluid to solid-like transition with increasing silica content, [57] as well as a high G' with respect to the viscous component G''.

From the master curve, the log-log slope at low frequencies which corresponds to the power law exponent was evaluated for each silica content and number of milling steps, and summarized in Figure 6a. The power law exponent of G' tends to decrease with increasing filler loading, presumably due to the formation of a hard silica NP network. As already discussed, with increasing milling, the slope of the power laws (in log-log presentation) of all composites increases due to the break-up of the physical crosslinks of the polymer network structure, thus fluidizing the composites. Interestingly, this effect is also observed in the unfilled compounds. It indicates that the unvulcanized rubber displays a naturally occurring network structure, which is softened by the mechanical action. It has been shown that such a network is built by the non-rubber components (proteins and phospholipids), which are known to induce long-chain branching and physical cross-linking in the rubber melt. [31, 58] Upon milling, this network structure is broken up by dispersing the physical crosslinks made of lipids and proteins, and this effect is not recovered by thermal treatment. The latter fact suggests that thermodynamics drives the system to a homogeneous mixture of NR and proteins, resp. phospholipids. As the silica network has been reconstituted leading to the same storage modulus by high-temperature treatment, the systematic evolution to more fluid-like rheology must be attributed to the irreversible break-up of the physical polymer crosslinks.

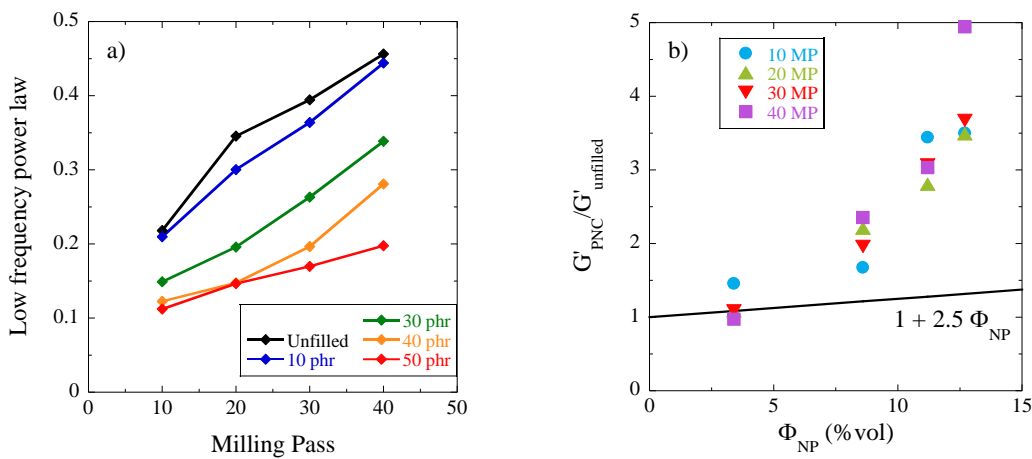

**Figure 6. a)** Slope of low frequency power law of G'(ω) as a function of milling passes and silica content. **(b)** Reinforcement factor from G' at 1000 rad/s normalized by the modulus of the corresponding unfilled compound (for each number of milling passes) as a function of the silica volume fraction. The line corresponds to purely hydrodynamic reinforcement (Einstein-Smallwood law [57]).

From the high-frequency elastic modulus (estimated here at $\omega$ = 1000 rad/s), one can determine the reinforcement factor, $G'_{PNC}/G'_{unfilled}$, with respect to the corresponding unfilled compound (prepared with the same number of milling passes). The rheological reinforcement factors are plotted in Figure 6b as a function of the silica content. In a polymer matrix, the addition of well-dispersed hard nanoparticles leads to an increase of the mechanical modulus also designated as the hydrodynamic reinforcement (see, e.g., [59, 60]). It is related to a perturbation of the strain field in the polymer and the measured stress is amplified. For spherical particles at low concentrations, it can be described by the Einstein-Smallwood [61] equation, $G'_{PNC}/G'_{unfilled} = 1 + 2.5\ \Phi_{NP}$, which assumes no particle interactions. One can see in Figure 6b that our data start to deviate from the theoretical prediction for silica loadings above 10 phr with a strong increase at the highest concentration corresponding to the formation of a large-scale silica structure possibly with percolated paths. Percolation at silica fractions in the range from 10 to 15%vol is commonly observed in PNCs. [51] Besides, there is no systematic impact of the number of milling passes on the reinforcement factors.

To summarize the rheological part of this study of the rejuvenation effect in such polymer nanocomposites, the evolution with milling is illustrated in Figure 7 in terms of the ratio of the low-strain storage modulus (cf. Figure 2a) to the one of the sample with 10 milling passes taken as reference. For samples which have not undergone a high-temperature thermal treatment (preparation at 70 °C), the ratio reproduces the strong decrease observed in Figure 2a with milling. The same ratio plotted for samples having undergone a high temperature treatment at 150 °C shows that the decrease is weakened, i.e. the effect of the milling partially cancelled, and the samples thus rejuvenated.

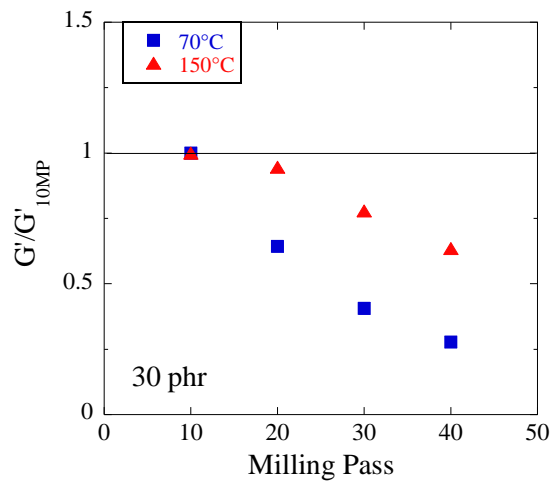

**Figure 7.** Ratio of the low-strain modulus G' with respect to the one of a reference having undergone 10 milling passes, as a function of milling, for two different thermal treatments before measurement at 70°C: equilibration at 70°C (squares) and annealing at 150°C for 5 minutes (triangles).

## 4. CONCLUSIONS

The impact of thermal treatment on the structural and dynamical properties of NR latex nanocomposites has been studied, using SAXS and rheology, as a function of silica content and milling. Milling is necessary in order to incorporate the silica filler, but the system is found to have structural and (non-linear) rheological characteristics which evolves with the amount of milling, from 10 to 40 passes in a two-roll mill. In absence of any thermal treatment, a strong strain-softening is observed in PNCs (Payne effect), the amplitude of which decreases with the number of milling passes: this is interpreted as silica network rupture, as it is not present in unfilled samples. The structure of these native samples has been studied by SAXS, and a clear incidence of the milling on the large-scale structure (> 100 nm) is found, thus corroborating the interpretation of the Payne effect in this system. At the same time, the local aggregate and primary particle structure remain unaffected by milling or higher concentrations, and are identical to the one of the initial powder. A quantitative analysis has been proposed of the scattering data in terms of the low-q limit of the inter-aggregate structure factor expressing the repulsiveness of the interactions. By comparing to the totally random description of polydisperse hard spheres describing the hard aggregates, it is found that after sufficient milling all data points collapse with the Percus-Yevick prediction, whereas with annealing they return to the more repulsive state they had initially, with a strong filler network structure.

In the mechanical study, this rejuvenation effect has also been observed after thermal treatment, allowing recovering a Payne effect which is considerably less dependent on milling. The term rejuvenation is meant to express the return at high temperature to the naturally aggregated state of the NPs, counterbalancing dispersion. In linear rheology, the storage modulus and the shape of G' and G" expressed by the low-frequency slopes are found to evolve from a more solid-like behavior to more liquid-like with milling (and the trend is opposite with thermal treatment), whereas the high-frequency storage modulus remains constant. The linear rheology study is consistent with the picture of a native structure of the silica, i.e. network formation, which forms spontaneously due to attractive silica-silica interactions, and absence of chemical coupling to the polymer. It is ruptured by milling, and reformed at high temperature. The strong increase of the reinforcement with silica content suggests the approach of percolation: the higher the silica volume fraction, the stronger the filler network, thus the higher the reinforcement of the linear shear modulus, the more gel-like the low frequency slopes, and the stronger the Payne effect. At the same time, underlying dynamics of the NR polymer is not modified neither by milling nor by the silica at the concentrations studied here, as shown by the VFT-analysis of the shift factors used to build the rheological master curve, and in agreement with the calorimetric $T_g$. Both milling and thermal treatment favor a more fluid-like flow behavior of the matrix, i.e. as opposed to the filler structure, there is no possibility of restructuration. The initial state of the matrix is thus never recovered, and rejuvenation only applies to the filler.

To summarize, our measurements of microscopic structure and dynamics of NR latex silica nanocomposites of both industrial and fundamental interest have provided a consistent picture of filler network formation and break-up as a function of filler content and milling, with direct consequences on macroscopic rheological properties. Driven by thermodynamics, the silica NPs tend to aggregate, and network formation is favored by thermal treatment after rupturing interparticle bonds, thereby rejuvenating the samples. The combination of milling and annealing may allow optimization of the rheological properties by fine-tuning the process of sample formation. Moreover, our study of the silica structure highlights the contributions of the large-scale structure to the rheology governed by the volume-spanning filler networks, whereas the underlying local aggregate structure formed by primary particles resists milling even at the highest silica volume fractions.

**Acknowledgements**. This work was financially supported by the Thailand Research Fund (TRF) through the Research and Researchers for Industries (RRi) and MBJ Enterprise Co. Ltd. (Grant no. PHD59I0048).